\documentclass[10pt, conference]{IEEEtran}

\usepackage[T1]{fontenc}
\usepackage[utf8]{inputenc}

\IEEEoverridecommandlockouts
\usepackage{cite}
\usepackage{amsmath,amssymb,amsfonts}
\usepackage{algorithmic}
\usepackage{graphicx}
\usepackage{textcomp}
\usepackage{xcolor}
\usepackage[table]{xcolor}
\usepackage{multirow}
\usepackage{xcolor}
\usepackage{xspace}
\usepackage{url}
\usepackage[most]{tcolorbox}
\usepackage{fontawesome5}
\usepackage{dcolumn}
\usepackage{booktabs}
\usepackage{graphicx}
\usepackage{tabularx}
\usepackage{pifont}
\usepackage{siunitx}
\newcommand{\sigC}{\textsuperscript{***}}
\newcommand{\sigB}{\textsuperscript{**}}
\newcommand{\sigA}{\textsuperscript{*}}
\newcolumntype{d}{D{.}{.}{-1}}
\newcolumntype{R}[1]{>{\raggedleft\arraybackslash}p{#1}}
\newcolumntype{L}[1]{>{\raggedright\arraybackslash}p{#1}}
\newcolumntype{C}{>{\centering\arraybackslash}X}
\definecolor{accentTRACE}{HTML}{1F4E79}
\definecolor{accentSPACE}{HTML}{1E6F5C}
\definecolor{accentObs}{HTML}{37474F}
\definecolor{coefPos}{HTML}{1F6F46}
\definecolor{coefNeg}{HTML}{C44E52}
\definecolor{coefNS}{HTML}{7A8693}
\newcommand{\icSat}{\textcolor{accentSPACE}{\faHeart}}
\newcommand{\icPerf}{\textcolor{accentSPACE}{\faChartLine}}
\newcommand{\icAct}{\textcolor{accentSPACE}{\faBolt}}
\newcommand{\icComm}{\textcolor{accentSPACE}{\faComments}}
\newcommand{\icEff}{\textcolor{accentSPACE}{\faStopwatch}}
\newcommand{\icT}{\textcolor{accentTRACE}{\faEye}}
\newcommand{\icR}{\textcolor{accentTRACE}{\faUserShield}}
\newcommand{\icA}{\textcolor{accentTRACE}{\faCopyright}}
\newcommand{\icC}{\textcolor{accentTRACE}{\faBan}}
\newcommand{\icE}{\textcolor{accentTRACE}{\faGavel}}

\newcommand{\cT}{\textcolor{accentTRACE}{Transparency}\xspace}
\newcommand{\cR}{\textcolor{accentTRACE}{Responsibility}\xspace}
\newcommand{\cA}{\textcolor{accentTRACE}{Attribution}\xspace}
\newcommand{\cC}{\textcolor{accentTRACE}{Constraints}\xspace}
\newcommand{\cE}{\textcolor{accentTRACE}{Enforcement}\xspace}
\newcommand{\cSat}{\textcolor{accentSPACE}{Satisfaction \& Well-Being}\xspace}
\newcommand{\cPerf}{\textcolor{accentSPACE}{Performance}\xspace}
\newcommand{\cAct}{\textcolor{accentSPACE}{Activity}\xspace}
\newcommand{\cComm}{\textcolor{accentSPACE}{Communication \& Collaboration}\xspace}
\newcommand{\cEff}{\textcolor{accentSPACE}{Efficiency \& Flow}\xspace}

\newcommand{\dT}{\icT\,\cT}
\newcommand{\dR}{\icR\,\cR}
\newcommand{\dA}{\icA\,\cA}
\newcommand{\dC}{\icC\,\cC}
\newcommand{\dE}{\icE\,\cE}
\newcommand{\dSat}{\icSat\,\cSat}
\newcommand{\dPerf}{\icPerf~\cPerf}
\newcommand{\dAct}{\icAct~\cAct}
\newcommand{\dComm}{\icComm~\cComm}
\newcommand{\dEff}{\icEff~\cEff}

\newtcolorbox{observation}[1]{
  enhanced, breakable, arc=3pt,
  colback=accentObs!6, colframe=accentObs, boxrule=0.8pt,
  before upper={\textbf{\textcolor{accentObs}{Observation~#1.}}\enspace},
  top=5pt, bottom=5pt, left=8pt, right=8pt,
  before skip=8pt, after skip=8pt,
}

\def\BibTeX{{\rm B\kern-.05em{\sc i\kern-.025em b}\kern-.08em
    T\kern-.1667em\lower.7ex\hbox{E}\kern-.125emX}}

\begin{document}

\title{Making AI Visible, Not Vanished: How AI Policies Reshape Developer Experience on GitHub}

\author{
\IEEEauthorblockN{Yunqi Chen}
\IEEEauthorblockA{\textit{University of California, Irvine}\\
USA\\
yunqic9@uci.edu}
\and
\IEEEauthorblockN{Thomas Zimmermann}
\IEEEauthorblockA{\textit{University of California, Irvine}\\
USA\\
tzimmer@uci.edu}
\and
\IEEEauthorblockN{Bianca Trinkenreich}
\IEEEauthorblockA{\textit{Colorado State University}\\
USA\\
bianca.trinkenreich@colostate.edu}
}

\maketitle

\begin{abstract}
Generative AI is rapidly reshaping Open Source Software (OSS) software development,prompting projects to introduce policies governing AI-assisted contributions. However, little is known about how these policies differ or whether they influence developer experience. We present the first large-scale empirical study of AI governance policies in OSS. Analyzing 29,624 GitHub repositories, we identify 385 projects that adopted AI policies and derive TRACE, a framework capturing five governance dimensions: Transparency, Responsibility, Attribution, Constraints, and Enforcement. We further classify policies into five governance families and estimate their effects using propensity-score matching and longitudinal difference-in-differences analysis. Our results show that AI governance primarily regulates rather than prohibits AI-assisted development. Policy adoption brings maintainer engagement, increased AI disclosure, richer review interactions, and improved code quality while AI-assisted contributions continue to grow. Governance design matters: policies emphasizing transparency and responsibility produced stronger community and quality outcomes than restrictive approaches alone. Our findings show how different AI governance strategies shape developer experience and provide evidence to help OSS communities design effective AI policies.

\end{abstract}

\section{Introduction}

Generative AI (GenAI) is transforming software development at an accelerating pace.
Developers now rely on AI-powered tools for code generation, documentation, testing,
and code review, while autonomous coding agents are beginning to navigate repositories,
generate patches, and participate directly in development workflows~\cite{peng2023impact,ziegler2024measuring}.
Although studies report short-term productivity gains, GenAI is also reshaping how
developers collaborate, learn, and reason about code, with consequences that extend
well beyond individual productivity~\cite{hou2024large}.

These changes carry particular weight for Open Source Software (OSS) communities.
Unlike traditional software organizations, OSS projects depend on decentralized
governance, volunteer contributions, and community-driven review processes to sustain
software quality and project health~\cite{eghbal2016roads}. As GenAI lowers the cost of
producing code, the volume of contributions communities must evaluate is increasing
rapidly, yet the effort required to review, validate, and integrate those contributions
remains largely human-driven. Recent work raises concerns about maintainer overload,
contributor accountability, code provenance, licensing uncertainty, and the long-term
sustainability of OSS communities in an era of AI-assisted development~\cite{alami2022pull,linaaker2024sustaining}.
Emerging evidence further suggests that AI-assisted contributions may introduce quality
risks, including increased complexity, technical debt, and verification burden, even
when short-term output rises~\cite{he2026speed}.

In response, OSS projects have begun establishing explicit policies governing AI-assisted contributions~\cite{yang2026beyond}. These policies address disclosure of AI usage~\cite{xiao2026self}, contributor accountability~\cite{alami2022pull}, acceptable forms of AI-generated
content~\cite{linaaker2024sustaining}, licensing and provenance
considerations~\cite{xu2025licoeval}, and interactions with autonomous
coding agents. Such policies represent a significant evolution of
OSS governance, extending existing community practices for managing contributions, maintaining quality, and sustaining collaborative work~\cite{alami2022pull,eghbal2016roads}. Yet, despite their growing prominence in practice, AI governance policies in OSS
remain poorly understood: their prevalence, diversity, and consequences have not been examined at scale.

Two critical research gaps remain open. First, while qualitative analyses have documented
isolated examples of AI governance practices, little is known about the \emph{prevalence, diversity, and structure of policy approaches} across the broader OSS ecosystem. Which governance models are emerging at scale? How do they differ? Second, and more
consequentially, the \emph{effects of these AI policies on communities} remain largely unexplored. OSS communities adopt governance mechanisms to influence contributor behavior and community outcomes, yet there is currently no empirical evidence on whether different AI policy approaches affect the developer experience: how developers perform, communicate,
collaborate, and sustain their participation over time.

To address both gaps, we present the first large-scale empirical study of AI governance
policies in OSS projects hosted on GitHub (Section~\ref{sec:design}). We mine repository governance artifacts,
including contribution guidelines, governance documents, and AI-specific policy files,
to identify and characterize the major families of AI governance policies emerging
across OSS communities (\textbf{RQ1}). We then investigate the consequences of these policy families on developer experience (\textbf{RQ2}). To estimate policy effects, we employ a longitudinal quasi-experimental design using Difference-in-Differences (DiD)
analysis on repository-level data spanning the period before and after policy adoption, enabling causal inference while accounting for pre-existing project-level trends.

This paper makes three contributions to advance the understanding of how AI governance is taking shape in practice and reshaping the lived experience of OSS communities.
\vspace{-0.5mm}
\begin{enumerate}
\item A large-scale characterization of AI governance policies in OSS, including dominant policy families emerging across GitHub projects (Section~\ref{sec:rq1}).
\item The first causal evidence on the effects of AI governance policies on developer experience, using longitudinal data and quasi-experimental methods (Section~\ref{sec:rq2}).
\item Actionable insights for OSS maintainers and community leaders seeking to balance the
opportunities of AI-assisted development with the challenge of sustaining healthy,
productive, and resilient OSS communities as well as a reusable framework called TRACE to guide future OSS AI research (Section~\ref{sec:insight}).
\end{enumerate}

\begin{table*}[t]
\centering
\caption{\textbf{TRACE dimension level definitions}}
\label{tab:trace_levels}
\small
\begin{sloppypar}\hyphenpenalty=10000\exhyphenpenalty=10000
\begin{tabular}{@{}p{25mm} >{\raggedright\arraybackslash}p{34mm} >{\raggedright\arraybackslash}p{34mm} >{\raggedright\arraybackslash}p{34mm} >{\raggedright\arraybackslash}p{34mm}@{}}
\toprule
Dimension & Level 1 & Level 2 & Level 3 & Level 4 \\
\midrule
\dT & No disclosure required. & Disclosure encouraged/ voluntary. & Basic disclosure (AI usage statement). & Detailed traceability (tool, prompt, transcript, files). \\[12pt]
\dR & No responsibility stated. & Human remains accountable. & Must explain/justify the work. & Must review/test/validate before submit. \\[12pt]
\dA & No license stated. & Generic IP/copyright stated. & Must satisfy license/DCO/ CLA compatibility. & Must document rights or provenance. \\[12pt]
\dC & No restriction stated. & Bulk/low-effort AI output restricted. & Restricted in specific contexts (issues, reviews). & Generally prohibited. \\[12pt]
\dE & No consequence stated. & Maintainer may request remediation/ignore. & Submission rejected/closed. & Future participation restricted/ban/suspension. \\
\bottomrule
\end{tabular}
\end{sloppypar}
\end{table*}

\section{Related Work}
\label{sec:related}

\textbf{GenAI Impact on Software Engineering and OSS communities.} GenAI is reshaping software engineering.
Recent literature has introduced methods to identify AI-assisted contributions in OSS, including heuristic-based approaches \cite{robbes2026promises}, LLM-based classifiers \cite{tufano2024unveiling, suh2025detecting}, self-reports \cite{xiao2026self}, and specilized neural classifiers \cite{daniotti2026who}.  Concurrently, research has evaluated  the impact of GenAI on  developer productivity, code quality \cite{ziegler2024measuring, he2026speed}, and code review \cite{branco2026lgtm, gao2026autopilot} They leverage diverse empirical methodologies, such as controlled and field experiments \cite{peng2023impact}, surveys \cite{liang2024survey}, randomized controlled trials (RCTs) \cite{becker2025measuringimpactearly2025ai, cui2026effects} and natural experiments \cite{song2026impactgenerativeaicollaborative,he2026speed}. While most studies agree on improvements in coding velocity, some point to risks in lower code quality and increased complexity \cite{he2026speed}. This influx has also triggered friction in OSS. Maintainers increasingly report overwhelmed by a surge of low-quality AI-generated submissions, often described in community discourse as ``AI slop'' \cite{stenberg2025slops,benson2026flood}.

\textbf{Governance in OSS communities.}
To manage the surge in AI activity, OSS communities are introduing related governance policies. Similar shifts have occurred in academia, where journals and conferences have adopted AI disclosure rules, though detected AI use there exceeds disclosed use by roughly 40 to 1~\cite{hebu2026journals}.
Traditionally, OSS communities govern development through Codes of Conduct \cite{sun2026beyond}, contribution guidelines \cite{coelho2017modern}, and pull request workflows \cite{alami2022pull}. Recent studies have begun treating these governance changes as interventions to measure their actual impact \cite{sun2026beyond}. Regarding GenAI specifically, recent qualtitive work has analyzed how major projects approach AI, classifying their practices into different governance orientations~\cite{yang2026beyond} and comparing self-reported AI use against official guidelines~\cite{xiao2026self}. We build on this governance-as-treatment perspective to evaluate how the adoption of AI policies affects the developer experience.

\section{Research Design}
\label{sec:design}

\begin{figure}[t]
\includegraphics[width=\columnwidth]{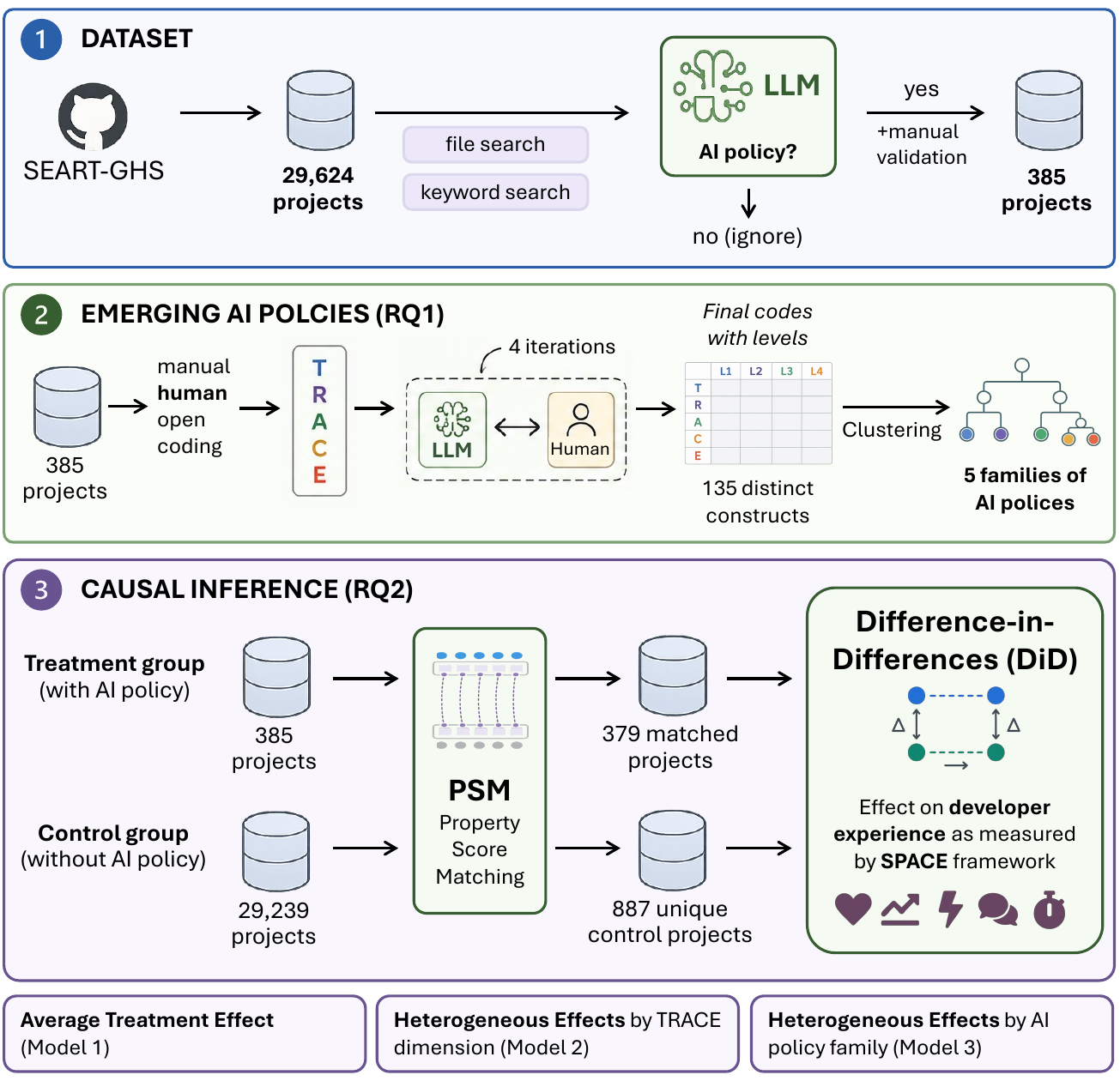}
\caption{Overview of the research design for this paper for \ding{202}\,identifying AI policies in open-source projects, \ding{203}\,deriving a taxonomy of policy constructs (RQ1), and \ding{204}\,estimating their effects on developer experience using matched controls and difference-in-differences models (RQ2).}
\label{fig:overview}
\end{figure}

Our research design has three parts (Fig.~\ref{fig:overview}). After the data collection, we first analyzed the OSS AI policy texts to identify governance dimensions they cover and to classify their overall regulatory posture. Next, we used causal inference to estimate the impact of these AI policies on different aspects of OSS developer experience. Our replication package is online\cite{anonymous_2026_21059330}.

\subsection{Data Collection and Preprocessing}
To build the dataset for this study, we used \textsc{SEART-GHS} \cite{dabic2021sampling}, a searchable dataset of \textmd{GitHub} projects, and filtered eligible repositories that:
\begin{itemize}
\item were created before January 2024, for sufficient development history.
\item accumulated at least 100 stars, indicating popularity.
\item contained over 100 commits from at least 10 distinct authors, indicating sustained activity.
\item had a valid license and was not a forked project.
\item maintained both open issues and pull requests (PRs), indicating active community engagement.
\item showed activity after January 2026, so the project was alive after the rise of coding agents.
\end{itemize}
Following this, we excluded non-software, educational, and toy repositories with keywords such as ``homework,'' ``course,'' and ``awesome'' \cite{zhou2019fork}. Projects that were deleted or made private were also removed.
Next, we used \textmd{GitHub} API to retrieve PR history for remaining projects.  We identified and flagged PRs from bots using a curated bot list \cite{Robbes2026AgenticMA} and GitHub user-type field.
Overall, the dataset comprised 29,624 projects, containing 10,192,203 PRs (median per project: 93), spanning from Jan 2025 to May 2026.

\subsection{Identifying AI Governance Policy Adoption}
To identify AI policies within the sample projects, we used GitHub code search API and repository tree scan to identify candidate AI policies, with filenames like ``AI\_POLICY.md'', ``AI\_USAGE\_POLICY.md''. Because some policies are embedded within broader contribution guidelines or governance documents, we also conducted text searches within documentation-like files using terms like ``AI policy'', ``AI usage policy'', ``AI contribution policy'' and used an LLM (\texttt{deepseek-v4-flash}) to classify each candidate as either a human-contributor-oriented policy, agent-guidance, or unrelated text. We then extracted and cleaned the policy text. We excluded purely agent-guidance files (e.g., \texttt{AGENTS.md}, \texttt{CLAUDE.md}) from our study.
We manually verified all candidate policies. If a file only contained a reference to an external policy file (e.g., \texttt{autobahn-js} \cite{autobahnjs} referenced \texttt{wamp-ai} \cite{wamp_ai_policy}), we retrieved the full text of that referenced policy.

We defined the \emph{adoption timestamp} of an AI policy as the merge time of the associated PR, or the commit timestamp if no PR was found. To ensure an 8-week pre-treatment and a 4-week post-treatment time window, we restricted the policy adoption period to between February 1, 2025, and April 30, 2026. Ultimately, we identified 385 treatment repositories that adopted an AI policy during this window.

\subsection{Characterizing AI Policies with the TRACE Framework}
\subsubsection {Deriving the TRACE framework}
We started our open coding process using an existing AI-governance framework~\cite{yang2026beyond}, adjusting formulation where better to capture the unique aspects of OSS AI policies. This process resulted in five dimensions: \dT (visibility of AI use), \dR (human accountability for AI output), \dA (source-rights and provenance), \dC (breadth of restriction on AI use), and \dE (consequences for violation), with definition of each level in Table~\ref{tab:trace_levels}. T, R, A, C, E are ordered from level 1 (weakest) to level 4 (strongest).

\subsubsection{Coding with LLMs and Iterative Human Validation}
After developing the TRACE framework, we used a large language model (LLM) (\texttt{deepseek-v4-flash}) to label each policy. We followed an iterative process starting with a codebook. In each iteration, we manually checked the LLM's coding results, updated the descriptions and boundary rules, and re-coded the sample cohort. We repeated this process until all level coding results matched our human judgment. Inter-rater reliability between one author and the LLM was assessed on a random 20\% sample ($n = 75$) of the 385-policy coding frame, reported by Cohen's Kappa value~\cite{landis1977measurement}.
Agreement was substantial across five TRACE dimensions: Transparency $\kappa = 0.79$, Responsibility $\kappa = 0.75$, Attribution $\kappa = 0.82$, Constraints $\kappa = 0.71$, and Enforcement $\kappa = 0.69$ (macro-mean $\kappa = 0.75$).

\subsection{Policy Family Taxonomy}
\label{subsec:families}
While the TRACE framework isolates individual dimensions, we also investigated their combined configurations. The coding results revealed that the 385 policies spanned 135 out of the 1,024 possible level combinations. To reduce this into a manageable set for analysis, we used a rule-based method to categorize the policies into five distinct families.

We first looked at how levels were distributed in each dimension: \dA is concentrated at its weakest level, \dR is clustered around the validation requirement, and \dT and \dE are bimodal distributions. Only \dC has a clear spread across all four levels, making it the best indicator of a project's governance stance toward AI. Based on this, we compared two classification methods: clustering approach (using Hamming distance) and a rule-based decision tree. The automatic clustering was flawed because its groups were driven almost entirely by a single dimension. In contrast, the rule-based method aligned directly with governance logic and created more balanced groups.

\subsection{Causal Inference Framework}
To estimate the causal effect of adopting an AI policy on the developer experience, we used \emph{difference-in-differences (DiD)}, a causal-inference method well-established in econometrics~\cite{card1994minimum,angrist2009mostly} and increasingly being applied in software engineering~\cite{fang2022tweets, song2026impactgenerativeaicollaborative, he2026speed, sun2026beyond, chen2026core}. DiD evaluates the effect of an intervention from observational data by comparing a \emph{treatment} group (repositories that adopted an AI policy) with a \emph{control} group (matched repositories that did not)~\cite{goodman2021difference}.

\begin{figure*}[t]
\centering
\includegraphics[width=\textwidth]{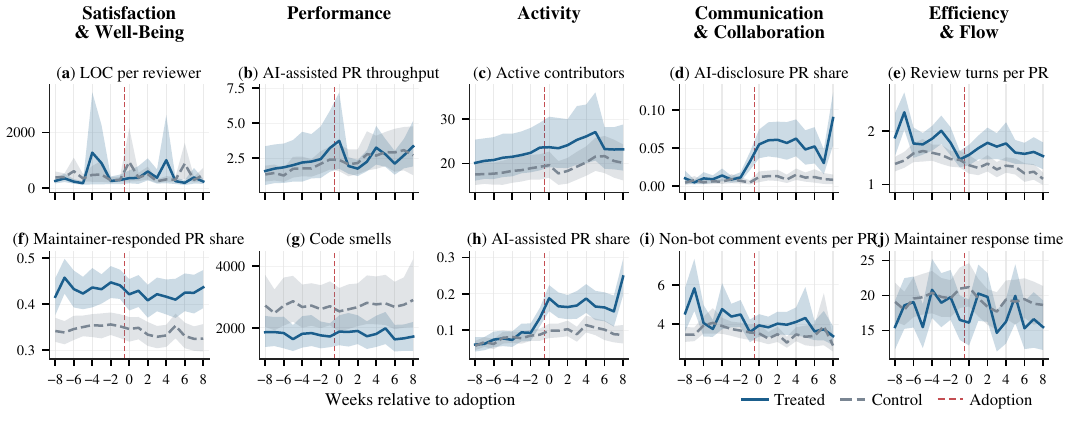}
\caption{ Event Study of 10 representative outcome variables (mean with 95\% CI), treated (solid) and matched control (dashed) repositories over ±8-week window. All outcomes successfully passed the parallel trends test within the pre-trend window, which is a prerequisite for the usage of difference-in-differences (DiD).}
\label{fig:mainevent}
\end{figure*}

The validity for DiD relies on the \emph{parallel trends} assumption, namely that, without the policy, treated and control repositories would have followed similar outcome trends over time. Since we cannot test this counterfactual directly, we supported it in two ways by (1) selecting control repositories using propensity score matching (PSM) based on \emph{pre-treatment} activity, and (2) testing the parallel trend with
event-study regressions on the pre-adoption period from week $-8$ to $-2$. All 19 outcomes successfully passed this parallel trends test within the pre-trend window, as shown in Fig.~\ref{fig:mainevent}. Because different repositories adopted AI policies at different times, we used a staggered DiD approach~\cite{callaway2021difference} that ``aligns'' each repository to its own adoption week (defined as event week 0) and groups all observations based on this relative time scale.

\subsection{Propensity Score Matching}
To construct a comparable control group, we used Propensity Score Matching (PSM)~\cite{rosenbaum1983central}, which matches each treated repository to never-treated ones as controls that shared similar probability of policy adoption. The unit of treatment is the repository at policy-adoption week $T_{i,0}$, with a pre-treatment window of Week~$-8$ to
Week~$-1$ (8 weeks). Week~$-1$ serves as the omitted event-study baseline while still included in propensity feature construction, and an older-history bucket aggregating activity before Week~$-8$. For each repository~$i$ in adoption cohort~$t$, the propensity feature $Z_{i,t}$ summarizes pre-treatment \emph{activity} extracted from \textsc{GHArchive}\cite{gharchive} event aggregates from Google BigQuery. We adopted a logistic regression model to calculate the propensity score. For repository~$i$, the feature vector~$Z_i$ captures the eight-week pre-treatment activity as below:
\begin{equation}
\begin{aligned}
\operatorname{logit}\Pr(D_{i}=1\mid Z_{i,t})
&= \alpha_t + \beta_{a}\ln a_{i,t-1} \\
&+ \sum_{k=1}^{8}\boldsymbol{\beta}'_{k}X_{i,-k}
   + \boldsymbol{\beta}'_{\mathrm{old}}X_{i,\leq-9},
\end{aligned}
\label{eq:psm}
\end{equation}

where $a_{i,t-1}$ is repository age at adoption, $X_{i,-k}$ counts features at week $i$ (e.g., PRs, stars, issues), $X_{i,\leq-9}$ aggregates the same components over all earlier weeks as an older-history baseline. We then matched each treated repository to up to three of its nearest never-treated controls by $k$-nearest-neighbor (KNN) matching on the estimated score. We restricted  candidates of the same main language and within a caliper of $0.2$ standard deviations~\cite{rosenbaum1985constructing}. Finally, we matched 379 of the 385 treated repositories to 887 unique controls.

\subsection{Outcome Variables: Developer Experience}

We conceptualized developer experience using the \textbf{SPACE} framework~\cite{forsgren2021space}, i.e., \dSat, \dPerf, \dAct, \dComm, and \dEff. We used SPACE as an organizing lens for 19 OSS outcome variables.
We computed PR-workflow outcomes through GitHub API to collect PR timeline events, reviews, and comments. Code-quality outcomes were obtained via a local SonarQube~\cite{sonarqube} snapshot scan of each repository at the last commit of each week. SonarQube successfully scanned 92\% (347 of 379) treated repositories and 87\% (768 of 887) of the matched controls, as the remaining failed to scan.

\smallskip
\dSat captures the human load around code review:
\textbf{Core-developer count} counts contributors who authored the top 80\% of the PRs within a 12-week rolling window.  \textbf{LOC per reviewer} divides the code churn (additions + deletions) by the number of unique, non-bot reviewers. \textbf{Maintainer-responded PR share} is the proportion of PRs with at least one review comment or timeline event from a maintainer.

\dPerf captures whether contributions get through and the quality of merged changes:
\textbf{AI-assisted PR throughput} counts the number of merged AI-assisted PRs. We identified AI-assisted PR with direct structural signals (agent branches or labels, co-author trailers, or self-disclosure)~\cite{Robbes2026AgenticMA}. Assuming developers continue to use AI tools after initial adoption, we labeled all subsequent PRs by the same author as AI-assisted. \textbf{Large-PR share} is the percentage of PRs opened with LOC $\geq$ 1{,}000, the GitHub XXL size band. The code-quality metrics come from weekly SonarQube scans: \textbf{duplicated-line density} (the percentage of duplicated lines), \textbf{vulnerabilities} (the count of security vulnerabilities SonarQube flags), \textbf{code smells} (the SonarSource maintainability-issue count), and \textbf{cognitive complexity} \cite{campbell2018cognitive}.

\dAct captures who contributes and how much of the contribution is AI-assisted:
\textbf{Active contributors} count non-bot users who participate in PR activities (open, merge, review, or any other actions in PR timelines). \textbf{AI-assisted PR share} is the number of AI-assisted PRs divided by the non-bot PRs opened. \textbf{Non-comment lines of code} (ncloc) is the weekly SonarQube count of non-blank, non-comment source lines.

\dComm captures the review process and the disclosure it relies on:
\textbf{AI-disclosure PR count} counts PRs containing AI self-disclosure phrases (``generated by,'' ``written by AI,'' ``AI-assisted,'' and similar phrases), detected via keyword-based matching on PR descriptions. \textbf{AI-disclosure PR share} is the count divided by the non-bot PRs opened. \textbf{Non-bot comment events per PR} is the average number of human comments and timeline events on a PR. \textbf{Bot first-response share} is the ratio of PRs with first review comment or timeline event from a bot account.

\dEff captures how readily review proceeds:
\textbf{Review turns per PR} is the mean number of alternating author--reviewer exchanges (discussion turns) in the PR comment thread. \textbf{Reviewers per PR} counts average unique non-bot reviewers who submit a review event on a PR. \textbf{Maintainer response time} measures median time from PR open to the first review comment or timeline event by a maintainer.

\subsection{Model Specification}

\subsubsection{Average Treatment Effect (Model I).}
For repository~$i$ in matched cohort~$g$ at calendar week~$t$, the average-effect model:
\begin{multline}
Y_{igt} = \beta\,(\mathrm{Treat}_i \times \mathrm{Post}_{it}) + \mathbf{X}_{it}'\boldsymbol{\gamma}\\
        + u_g + \lambda_t + \varepsilon_{igt},
\label{eq:did}
\end{multline}
where $Y_{igt}$ is the outcome (log-transformed for count-based metrics). $\mathrm{Treat}_i$ marks repositories that adopted an AI policy and $\mathrm{Post}_{it}$ marks calendar weeks after $i$'s adoption. $\mathbf{X}_{it}$ is the vector of weekly GHArchive control variables (repository age and the weekly counts of opened issues, new stars, and distinct users). $u_g$ is a matched-set random intercept, where a treated repository and its matched controls share one identifier, and $\lambda_t$ is a calendar-week random intercept.

Our main coefficient of interest is $\beta$, which estimates average treatment effect (ATT) on the treated,  relative to counterfactual trend of their matched controls. For log-transformed outcomes $\beta$, we report the exponential retransformation $(e^{\beta}-1)\times100$. For share and rate outcomes, we report effect as an absolute percentage change on the original scale.

\subsubsection{Heterogeneous effects by TRACE dimension (Model II) and policy family (Model III).}
To evaluate  whether policy impacts vary based on their specific content, we extend Eq.~\eqref{eq:did} by interacting the treatment effect with a policy attribute $M_i$:
\begin{multline}
Y_{igt} = \alpha\,(\mathrm{Treat}_i \times \mathrm{Post}_{it})
        + \boldsymbol{\beta}\,\bigl(\mathrm{Treat}_i \times \mathrm{Post}_{it}\times M_i\bigr)\\
        + \mathbf{X}_{it}'\boldsymbol{\gamma} + u_g + \lambda_t + \varepsilon_{igt},
\label{eq:ddd}
\end{multline}
where $M_i$ is a TRACE dimension level or the policy family. For a focal dimension, the model includes treated-post interactions for all five TRACE dimensions, but reports the focal dimension's level-2--4 differentials against level~1, so the other four dimensions are held at their reference levels. For families, the model includes all 5 family indicators simultaneously with Permissive \& Open as the reference. The each family ATT is reconstructed by summing the baseline effect $\alpha$ and its corresponding interaction term. The vector $\boldsymbol{\beta}$ thus captures how much the baseline policy effect is moderated by the specific governance posture.

We report the results using an 8-weeks post-treatment window. Robustness checks using alternative windows of 4, 6, 10, and 12 weeks yielded consistent directional results. Model fit is evaluated using marginal and conditional $R^2$ ($R^2_m$, $R^2_c$) metrics~\cite{nakagawa2013general, johnson2014extension}. Details are in the replication package \cite{anonymous_2026_21059330}.

\begin{figure}[t]
  \centering
  \includegraphics[width=\columnwidth]{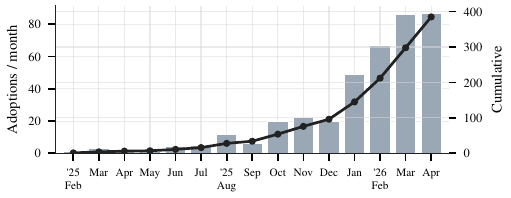}
\caption{Monthly/Cumulative adoption trend of AI policies}
  \label{fig:adoption}
\end{figure}
\begin{figure}[t]
  \centering
  \includegraphics[width=\columnwidth]{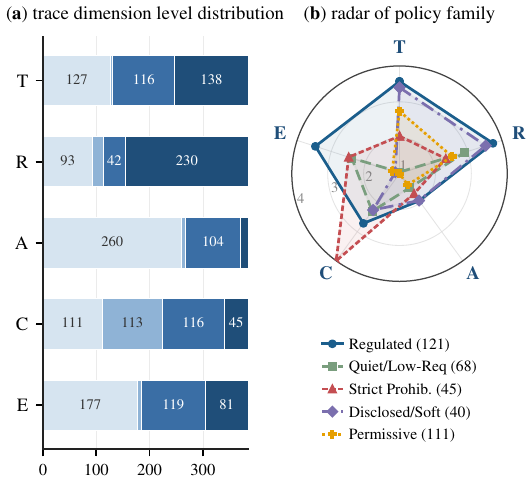}
\caption{(a) TRACE dimension level distribution across the 385 policies. (b) Radar Overlay of Average TRACE profile for each policy family.}
  \label{fig:traceradar}
\end{figure}

\section{Prevalence and Characterization of OSS AI Governance Policies (RQ1)}
\label{sec:rq1}

Among the 29,624 sample projects, we discovered 385 repositories that adopted AI policy within our time window. As shown in Fig.~\ref {fig:adoption}, the adoption was sparse through 2025 and accelerated in 2026. Only one policy was adopted in February 2025, compared to 49 in January, 67 in February, 86 in March, and 87 in April 2026.

\subsection{TRACE Dimension Distribution}

The distribution of TRACE levels is uneven across five dimensions (see Fig.~\ref{fig:traceradar}(a)), indicating that OSS AI governance does not align along a single permissive-to-restrictive axis. We use excerpts from AI policies to exemplify each dimension.
\smallskip

\dT. 33.0\% of policies (127/385) impose no disclosure requirement (T1). At T2, 1.0\% (4/385) encourage disclosure or optional, as in \texttt{Promptfoo} (``disclosure of AI usage is optional.'') At T3, 30.1\% (116/385) require basic AI usage disclosure like \texttt{Coder} stating: ``Contributors must disclose AI involvement in the pull request description whenever these guidelines apply'' At T4, 35.8\% (138/385) demand detailed traceability, such as \texttt{Ghostty}'s  requirement: ``All AI usage in any form must be disclosed. You must state the tool you used along with the extent that the work was AI-assisted.''

\dR. 24.2\% of policies (93/385) state no AI-specific responsibility (R1). At R2, 5.2\% (20/385) demand human authorship, as \texttt{libCEED} states: ``The human creating the PR is ultimately responsible for the content in the PR.'' At R3, 10.9\% (42/385) require contributors to understand and explain AI output, as \texttt{AnkiDroid} insists: ``You must be able to explain all your contributions.'' At R4, 59.7\% (230/385) mandate human review and validation, as \texttt{vLLM} insists: ``Review and understand AI-generated code with the same care as code you write manually.''

\dA. 67.5\% of policies (260/385) do not address attribution (A1). At A2, 1.8\% (7/385) assert general IP responsibility, as \texttt{DSharpPlus} states: ``We require that you own the intellectual property to any contribution you choose to submit.'' At A3, 27.0\% (104/385) require license or source compatibility, as \texttt{Envoy} rules: ``All generated code must be released under the same license as Envoy. You are responsible for ensuring that the tools you use to generate code do not add any additional licensing restrictions.'' At A4, 3.6\% (14/385) demand documented rights or provenance, as \texttt{AxoSyslog} requires, with the submitter ``adding their own Signed-off-by tag to certify the DCO.''

\dC. 28.8\% of policies (111/385) impose no constraints (C1). At C2, 29.4\% (113/385) decline full or bulk AI submissions, as \texttt{detekt} notes ``a special policy for contributions that appear to be entirely generated by AI,'' which covers ``PRs that we suspect have little to no human scrutiny or review.'' At C3, 30.1\% (116/385) restrict AI use by specific context or channel, as \texttt{Directus} rules: ``Do not copy-paste or use entirely AI-generated text for issue descriptions, pull requests, comments, or replies.'' At C4, 11.7\% (45/385) completely ban AI submission: ``AI-generated content is not allowed to be submitted to the repository'' (\texttt{Delta-V}).

\dE. 46.0\% of policies (177/385) omit consequence for violation (E1). At E2, 2.1\% (8/385) specify remediation or review limitation,as \texttt{openEMS} warns: ``Reviewers will apply extra scrutiny to \texttt{Generated-by} submissions, and overly long AI-generated descriptions will be asked to be trimmed before review.''. At E3, 30.9\% (119/385) state that current submissions will be rejected or closed, as \texttt{Apache Arrow} warns: ``Rs that appear to be fully generated by AI with little to no engagement from the author may be closed without further review.'' At E4, 21.0\% (81/385) restrict future participation, as \texttt{Open Shading Language} warns that ``the account associated with the submission may be banned from future project participation.''

\begin{observation}{1} Most policies require AI disclosure (T3--T4, 66.0\%) and mandate human validation responsibility (R4, 59.7\%), while license attrition remains largely absent (A1, 67.5\%). Constraints are common (71.2\% above C1), but enforcement is split between no sanction, rejection, or banning future participation. \end{observation}

\subsection{Policy Families}

As explained in Section`\ref{subsec:families}, we used a rule-based approach to group policies with similar level combinations into families.
\dC defines the two extremes: \textbf{Strict Prohibition} ($C = 4$) and \textbf{Permissive \& Open} ($C = 1$). The conditional-use middle group ($C \in \{2, 3\}$) is first split by disclosure requirements (\dT) and then by enforcement (\dE). This resulted in three families: \textbf{Quiet \& Low-Requirement} (no disclosure, $T < 3$), \textbf{Disclosed but Soft-Enforced} (requires disclosure but weak enforcement, $T \geq 3, E < 3$), and \textbf{Regulated \& Controlled} (requires disclosure and strict enforcement, $T \geq 3, E \geq 3$).
The TRACE profile for each policy family is shown in Fig.~\ref{fig:traceradar}(b).

\medskip

\textbf{Permissive \& Open  (111, 28.8\%).} Defined by C1 (no restriction), this family imposes little to no AI-specific governance. These policies typically treat AI tools as ordinary development aids and apply the standard review criteria to all submissions without additional mandates. For example, \texttt{sos} states that it ``welcomes contributions made with the assistance of AI coding assistants and agents as long as they meet all the normal criteria for correctness,'' with no disclosure mandate or AI-specific restriction.

\textbf{Disclosed but Soft-Enforced (40, 10.4\%).} Defined by moderate restriction (C2 or C3) with disclosure requirement (T$\geq$3) but only mild enforcement (E<3), this family renders AI use visible without strong sanctions. \texttt{vLLM} requires contributors to disclose AI assistance in PR descriptions and attribute AI tools via commit trailers such as \texttt{Co-authored-by}, while specifying no consequences for non-disclosure beyond the review standards that apply to all code.

\textbf{Regulated \& Controlled (121, 31.4\%).} The largest family, defined by moderate restriction (C2 or C3), disclosure requirement (T$\geq$3), and strong enforcement (E$\geq$3, it mandates visibility and backs it with explicit consequences. For example, \texttt{Ghostty} requires that all AI usage be disclosed with the specific tool named, restricts AI-generated PRs to accepted issues only, and closes any PR where AI use is suspected but undisclosed.

\textbf{Quiet \& Low-Requirement (68, 17.7\%).} Defined by moderate restriction (C2 or C3) without a disclosure mandate (T<3), this family imposes constraints on how AI may be used but does not require contributors to declare AI use. \texttt{simdjson} requires a ``human in the loop'' and holds contributors fully accountable for AI-assisted work, expecting them to answer review questions and start with small contributions, yet does not mandate disclosure of specific tools used.

\textbf{Strict Prohibition (45, 11.7\%).} Defined by C4 (generally excluded), this family bans AI-generated contributions outright or permits only narrow assistive use.\texttt{OctoPrint} declares that it ``does not accept pull requests that are fully or predominantly AI-generated,'' restricting AI tools to ``an assistive capacity'' where ``the majority of the code is authored by a human contributor.''

\begin{observation}{2}
OSS Policy families show governed permission rather than prohibition-centered. Regulated \& Controlled is the largest family (121, 31.4\%), followed by Permissive \& Open (111, 28.8\%). Quiet \& Low-Requirement policies form a middle category (68, 17.7\%). Disclosed but Soft-Enforced (40, 10.4\%) and Strict Prohibition policies are minority (45, 11.7\%).
\end{observation}

\section{Effects of AI Governance Policies on OSS Developer Experience (RQ2)}
\label{sec:rq2}

\subsection{Average Treatment Effects on Developer Experience}
\begin{table}[!t]
\centering
\caption{\textbf{Model I: Average impact of AI policy adoption on 19 outcomes within the 8-week post-adoption window.}}
\label{tab:maineffects}
\small
\renewcommand{\arraystretch}{1.12}
\setlength{\tabcolsep}{4pt}
\begin{tabular}{@{}>{\raggedright\arraybackslash}p{42mm}R{4mm}@{}L{6.5mm}>{\centering\arraybackslash}p{18mm}>{\centering\arraybackslash}p{9mm}@{}}
\toprule
Outcome & \multicolumn{2}{c}{$\beta$} & $R^2_m\,(R^2_c)$ & N.Obs. \\
\midrule
\multicolumn{5}{@{}l}{\textcolor{accentSPACE}{\dSat}} \\
\midrule
Core developer count (log) & \textcolor{coefPos}{.} & \textcolor{coefPos}{11\sigC} & 0.44\,(\,0.56) & 20{,}623 \\
LOC per reviewer (log) & \textcolor{coefNeg}{-.} & \textcolor{coefNeg}{16\sigC} & 0.03\,(\,0.12) & 11{,}116 \\
Maintainer responded PR share & \textcolor{coefPos}{.} & \textcolor{coefPos}{06\sigC} & 0.03\,(\,0.20) & 15{,}485 \\
\midrule
\multicolumn{5}{@{}l}{\textcolor{accentSPACE}{\dPerf}} \\
\midrule
AI-assisted PR throughput (log) & \textcolor{coefPos}{.} & \textcolor{coefPos}{10\sigC} & 0.19\,(\,0.35) & 20{,}623 \\
Large PR share & \textcolor{coefNeg}{-.} & \textcolor{coefNeg}{02\sigC} & 0.02\,(\,0.08) & 15{,}485 \\
Duplicated lines density & \textcolor{coefNeg}{-1.} & \textcolor{coefNeg}{69\sigC} & 0.01\,(\,0.32) & 13{,}976 \\
Vulnerabilities (log) & \textcolor{coefNeg}{-.} & \textcolor{coefNeg}{11\sigC} & 0.02\,(\,0.43) & 13{,}976 \\
Code smells (log) & \textcolor{coefNeg}{-.} & \textcolor{coefNeg}{17\sigC} & 0.05\,(\,0.60) & 13{,}976 \\
Cognitive complexity (log) & \textcolor{coefNeg}{-.} & \textcolor{coefNeg}{18\sigC} & 0.05\,(\,0.68) & 13{,}976 \\
\midrule
\multicolumn{5}{@{}l}{\textcolor{accentSPACE}{\dAct}} \\
\midrule
Active contributors (log) & \textcolor{coefPos}{.} & \textcolor{coefPos}{12\sigC} & 0.62\,(\,0.71) & 20{,}623 \\
AI-assisted PR share & \textcolor{coefPos}{.} & \textcolor{coefPos}{07\sigC} & 0.04\,(\,0.22) & 16{,}372 \\
Lines of code (ncloc) (log) & \textcolor{coefNeg}{-.} & \textcolor{coefNeg}{22\sigC} & 0.05\,(\,0.68) & 13{,}976 \\
\midrule
\multicolumn{5}{@{}l}{\textcolor{accentSPACE}{\dComm}} \\
\midrule
AI disclosure PR count (log) & \textcolor{coefPos}{.} & \textcolor{coefPos}{11\sigC} & 0.10\,(\,0.23) & 20{,}623 \\
AI disclosure PR share & \textcolor{coefPos}{.} & \textcolor{coefPos}{04\sigC} & 0.03\,(\,0.12) & 16{,}372 \\
Non-bot comments/PR (log) & \textcolor{coefPos}{.} & \textcolor{coefPos}{08\sigC} & 0.11\,(\,0.23) & 15{,}312 \\
Bot first response share & \textcolor{coefNeg}{-.} & \textcolor{coefNeg}{03\sigC} & 0.04\,(\,0.32) & 14{,}006 \\
\midrule
\multicolumn{5}{@{}l}{\textcolor{accentSPACE}{\dEff}} \\
\midrule
Review turns/PR (log) & \textcolor{coefPos}{.} & \textcolor{coefPos}{08\sigC} & 0.12\,(\,0.25) & 15{,}485 \\
Reviewers/PR (log) & \textcolor{coefPos}{.} & \textcolor{coefPos}{04\sigC} & 0.11\,(\,0.25) & 15{,}485 \\
Maintainer response time (log) & \textcolor{coefNeg}{-.} & \textcolor{coefNeg}{08\sigA} & 0.02\,(\,0.11) & 11{,}380 \\
\bottomrule
\multicolumn{5}{p{0.92\columnwidth}}{\footnotesize $^{***}p<0.001$, $^{**}p<0.01$, $^{*}p<0.05$.}\\
\end{tabular}
\end{table}

\begin{table*}[!t]
\centering
\caption{\textbf{Model II: Heterogeneous effects by TRACE dimension}}
\label{tab:ddd_dim}
\small
\setlength{\tabcolsep}{5pt}
\renewcommand{\arraystretch}{1.0}
\begin{tabular}{@{}l *{10}{c}@{}}
\toprule
Outcome & \multicolumn{2}{c}{\icT\,\textcolor{accentTRACE}{\textbf{Transparency}}} & \multicolumn{2}{c}{\icR\,\textcolor{accentTRACE}{\textbf{Responsibility}}} & \multicolumn{1}{c}{\icA\,\textcolor{accentTRACE}{\textbf{Attribution}}} & \multicolumn{3}{c}{\icC\,\textcolor{accentTRACE}{\textbf{Communication}}} & \multicolumn{2}{c}{\icE\,\textcolor{accentTRACE}{\textbf{Enforcement}}} \\
\cmidrule(lr){2-3}\cmidrule(lr){4-5}\cmidrule(lr){6-6}\cmidrule(lr){7-9}\cmidrule(lr){10-11}
 & L3 & L4 & L3 & L4 & L3 & L2 & L3 & L4 & L3 & L4 \\
\midrule
\multicolumn{11}{@{}l}{\textcolor{accentSPACE}{\textbf{\icSat~Satisfaction}}} \\
\midrule
Core developer count (log) & \textcolor{coefPos}{.16\sigB} & \textcolor{coefPos}{.16\sigB} & \textcolor{coefNS}{--} & \textcolor{coefPos}{.13\sigA} & \textcolor{coefNeg}{-.05\sigC} & \textcolor{coefNS}{--} & \textcolor{coefNeg}{-.16\sigC} & \textcolor{coefNS}{--} & \textcolor{coefNS}{--} & \textcolor{coefNS}{--} \\
LOC per reviewer (log) & \textcolor{coefNS}{--} & \textcolor{coefNS}{--} & \textcolor{coefNS}{--} & \textcolor{coefNS}{--} & \textcolor{coefNS}{--} & \textcolor{coefNS}{--} & \textcolor{coefNS}{--} & \textcolor{coefNS}{--} & \textcolor{coefNS}{--} & \textcolor{coefNS}{--} \\
Maintainer responded PR share & \textcolor{coefNS}{--} & \textcolor{coefPos}{.08\sigB} & \textcolor{coefNeg}{-.03\sigA} & \textcolor{coefNeg}{-.04\sigC} & \textcolor{coefNS}{--} & \textcolor{coefPos}{.07\sigA} & \textcolor{coefNS}{--} & \textcolor{coefNS}{--} & \textcolor{coefNS}{--} & \textcolor{coefPos}{.09\sigB} \\
\midrule
\multicolumn{11}{@{}l}{\textcolor{accentSPACE}{\textbf{\icPerf~Performance}}} \\
\midrule
AI-assisted PR throughput (log) & \textcolor{coefNS}{--} & \textcolor{coefPos}{.34\sigC} & \textcolor{coefPos}{.36\sigC} & \textcolor{coefPos}{.25\sigB} & \textcolor{coefNS}{--} & \textcolor{coefNeg}{-.18\sigC} & \textcolor{coefNeg}{-.40\sigC} & \textcolor{coefNeg}{-.24\sigC} & \textcolor{coefNS}{--} & \textcolor{coefPos}{.22\sigA} \\
Large PR share & \textcolor{coefNS}{--} & \textcolor{coefNS}{--} & \textcolor{coefNS}{--} & \textcolor{coefNS}{--} & \textcolor{coefPos}{.01\sigA} & \textcolor{coefNS}{--} & \textcolor{coefNS}{--} & \textcolor{coefNS}{--} & \textcolor{coefPos}{.01\sigB} & \textcolor{coefPos}{.02\sigB} \\
Duplicated lines density & \textcolor{coefNS}{--} & \textcolor{coefNS}{--} & \textcolor{coefNS}{--} & \textcolor{coefNS}{--} & \textcolor{coefNS}{--} & \textcolor{coefNS}{--} & \textcolor{coefNS}{--} & \textcolor{coefNS}{--} & \textcolor{coefNS}{--} & \textcolor{coefNS}{--} \\
Vulnerabilities (log) & \textcolor{coefNS}{--} & \textcolor{coefNS}{--} & \textcolor{coefNS}{--} & \textcolor{coefNS}{--} & \textcolor{coefPos}{.20\sigB} & \textcolor{coefNS}{--} & \textcolor{coefNS}{--} & \textcolor{coefNS}{--} & \textcolor{coefNeg}{-.09\sigA} & \textcolor{coefNeg}{-.16\sigA} \\
Code smells (log) & \textcolor{coefPos}{.39\sigC} & \textcolor{coefPos}{.01\sigA} & \textcolor{coefNS}{--} & \textcolor{coefNeg}{-.61\sigB} & \textcolor{coefNeg}{-.61\sigC} & \textcolor{coefNeg}{-.01\sigA} & \textcolor{coefNS}{--} & \textcolor{coefNS}{--} & \textcolor{coefNS}{--} & \textcolor{coefNS}{--} \\
Cognitive complexity (log) & \textcolor{coefPos}{.27\sigC} & \textcolor{coefPos}{.03\sigC} & \textcolor{coefPos}{.02\sigB} & \textcolor{coefNeg}{-.64\sigA} & \textcolor{coefNeg}{-.76\sigC} & \textcolor{coefNS}{--} & \textcolor{coefNS}{--} & \textcolor{coefNS}{--} & \textcolor{coefNS}{--} & \textcolor{coefNS}{--} \\
\midrule
\multicolumn{11}{@{}l}{\textcolor{accentSPACE}{\textbf{\icAct~Activity}}} \\
\midrule
Active contributors (log) & \textcolor{coefPos}{.15\sigC} & \textcolor{coefPos}{.17\sigC} & \textcolor{coefNS}{--} & \textcolor{coefNS}{--} & \textcolor{coefNeg}{-.09\sigC} & \textcolor{coefPos}{.13\sigB} & \textcolor{coefNeg}{-.08\sigA} & \textcolor{coefNS}{--} & \textcolor{coefNS}{--} & \textcolor{coefNeg}{-.10\sigB} \\
AI-assisted PR share & \textcolor{coefPos}{.11\sigC} & \textcolor{coefPos}{.15\sigC} & \textcolor{coefNS}{--} & \textcolor{coefNS}{--} & \textcolor{coefPos}{.09\sigB} & \textcolor{coefNeg}{-.02\sigC} & \textcolor{coefNeg}{-.09\sigC} & \textcolor{coefNeg}{-.04\sigC} & \textcolor{coefNS}{--} & \textcolor{coefPos}{.10\sigB} \\
Lines of code (ncloc) (log) & \textcolor{coefPos}{.42\sigC} & \textcolor{coefPos}{.22\sigB} & \textcolor{coefNS}{--} & \textcolor{coefNeg}{-.47\sigB} & \textcolor{coefNeg}{-.58\sigC} & \textcolor{coefNS}{--} & \textcolor{coefNS}{--} & \textcolor{coefNS}{--} & \textcolor{coefNeg}{-.40\sigA} & \textcolor{coefNS}{--} \\
\midrule
\multicolumn{11}{@{}l}{\textcolor{accentSPACE}{\textbf{\icComm~Communication}}} \\
\midrule
AI disclosure PR count (log) & \textcolor{coefPos}{.12\sigC} & \textcolor{coefPos}{.21\sigC} & \textcolor{coefPos}{.11\sigC} & \textcolor{coefPos}{.07\sigC} & \textcolor{coefPos}{.05\sigB} & \textcolor{coefNeg}{-.05\sigB} & \textcolor{coefNeg}{-.19\sigC} & \textcolor{coefNeg}{-.11\sigC} & \textcolor{coefPos}{.06\sigB} & \textcolor{coefNS}{--} \\
AI disclosure PR share & \textcolor{coefPos}{.05\sigC} & \textcolor{coefPos}{.04\sigC} & \textcolor{coefPos}{.02\sigB} & \textcolor{coefPos}{.02\sigC} & \textcolor{coefPos}{.01\sigC} & \textcolor{coefNS}{--} & \textcolor{coefNeg}{-.08\sigC} & \textcolor{coefNeg}{-.03\sigC} & \textcolor{coefPos}{.04\sigC} & \textcolor{coefNS}{--} \\
Non-bot comments/PR (log) & \textcolor{coefNS}{--} & \textcolor{coefNS}{--} & \textcolor{coefNS}{--} & \textcolor{coefNS}{--} & \textcolor{coefPos}{.76\sigA} & \textcolor{coefNS}{--} & \textcolor{coefNS}{--} & \textcolor{coefNS}{--} & \textcolor{coefNS}{--} & \textcolor{coefNS}{--} \\
Bot first response share & \textcolor{coefNeg}{-.08\sigA} & \textcolor{coefNS}{--} & \textcolor{coefNS}{--} & \textcolor{coefPos}{.09\sigC} & \textcolor{coefNS}{--} & \textcolor{coefNS}{--} & \textcolor{coefNS}{--} & \textcolor{coefNS}{--} & \textcolor{coefNeg}{-.12\sigC} & \textcolor{coefNS}{--} \\
\midrule
\multicolumn{11}{@{}l}{\textcolor{accentSPACE}{\textbf{\icEff~Efficiency}}} \\
\midrule
Review turns/PR (log) & \textcolor{coefNS}{--} & \textcolor{coefNS}{--} & \textcolor{coefNS}{--} & \textcolor{coefNS}{--} & \textcolor{coefNS}{--} & \textcolor{coefNS}{--} & \textcolor{coefNS}{--} & \textcolor{coefNS}{--} & \textcolor{coefNS}{--} & \textcolor{coefNS}{--} \\
Reviewers/PR (log) & \textcolor{coefNS}{--} & \textcolor{coefNS}{--} & \textcolor{coefPos}{.03\sigB} & \textcolor{coefPos}{.05\sigC} & \textcolor{coefNeg}{-.00\sigB} & \textcolor{coefNS}{--} & \textcolor{coefNS}{--} & \textcolor{coefNeg}{-.12\sigA} & \textcolor{coefNS}{--} & \textcolor{coefNS}{--} \\
Maintainer response time (log) & \textcolor{coefPos}{.02\sigA} & \textcolor{coefNS}{--} & \textcolor{coefNS}{--} & \textcolor{coefPos}{.04\sigA} & \textcolor{coefNeg}{-.38\sigA} & \textcolor{coefNS}{--} & \textcolor{coefNS}{--} & \textcolor{coefNS}{--} & \textcolor{coefNS}{--} & \textcolor{coefNS}{--} \\
\bottomrule
\end{tabular}
\par\medskip
Each cell is the ATT coefficient $\beta$ for the focal TRACE level with the other dimensions held at level~1; \\``\textcolor{coefNS}{--}'' marks $p\ge0.05$ and levels with $\leq$ 20  treated policies. $^{***}p<0.001$, $^{**}p<0.01$, $^{*}p<0.05$.
\end{table*}

Table~\ref{tab:maineffects} reports the estimated effects of AI policy adoption on the 19 outcome variables within the 8-week post-adoption window.
\smallskip

\dSat. Maintainer responded ratio rises 6\% ($\beta = .06^{***}$) from 44\% to about half of pull requests. Core developer count rises 12\% ($\beta = .11^{***}$), from about 6 to 7. LOC per reviewer decreases 14\% ($\beta = -.16^{***}$), from about 495 to about 425 lines per reviewer.

\dPerf. AI-assisted PR throughput rises 10\% ($\beta = .10^{***}$), while the large-PR ratio ($> 1000$ LOC)  falls 2.1\% ($\beta = -.02^{***}$), from 10.3\% to about 8.1\%. Regarding SonarQube-based quality metrics, vulnerabilities decline 11\% ($\beta = -.11^{***}$) from about 3.7 to about 3.3 per repository, duplicated-line density declines about 1.69 ($\beta = -1.69^{***}$) from 5.7 to about 4.0, code smells fall 16\% ($\beta = -.17^{***}$), and cognitive complexity falls 17\% ($\beta = -.18^{***}$).

\dAct. Active contributors increase 13\% ($\beta = .12^{***}$), meaning 2 additional contributors per week. AI-assisted PR share rises 6.8\% ($\beta = .07^{***}$), from a pre-adoption level of 6.4\% to about 13\% of pull requests, nearly doubling. AI-assisted merged throughput rises 10\% ($\beta = .10^{***}$), though the absolute gain is modest (1.9 to 2.1). Lines of code fall 20\% ($\beta = -.22^{***}$).

\dComm. PRs disclosing AI use rise 11\% ($\beta = .11^{***}$), and disclosure share rises 4.2\% ($\beta = .04^{***}$), from below 1\% to about 5\% average per repository. The 11\% count gain is about 1 extra disclosure PR every 10 repo-weeks. Non-bot comments per PR rise about 8\% ($\beta = .08^{***}$), while the bot first-response share decreases 3\%  ($\beta = -.04^{***}$)from a baseline of 37\%.

\dEff.  Review turns per PR increase about 8\% ($\beta = .08^{***}$), reviewers per PR rise 4\% ($\beta = .04^{***}$), and maintainer response time falls about 8\% ($\beta = -.08^{*}$) from about 18 to about 17 hours.

\begin{observation}{3}
After AI policy adoption, projects exhibit:
(1) higher human participation, maintainer involvement and review interaction (maintainer-responded share +6\%, bot first-response share \mbox{--3\%}, LOC per reviewer \mbox{--14\%}, review turns +8\%, reviewers +4\%);
(2) increased AI tool usage coupled with greater transparency
(AI-assisted PR share +6.8\%, disclosure share +4.2\%, AI-assisted PR throughput +10\%);
and (3) improved code quality  (code smells \mbox{--16\%}, cognitive complexity \mbox{--17\%}).
\end{observation}

\subsection{Heterogeneous Effects by TRACE Dimension}

Table~\ref{tab:ddd_dim} reports heterogeneous treatment effects by TRACE dimension. Each cell is the ATT when the focal dimension is raised to that level, with the others held at level~1. We report only levels supported by over 20 treated policies.

\subsubsection*{\dT~amplifies disclosure, AI-assisted activity, and community participation}
 Higher Transparency levels correspond to larger gains in disclosure and AI-assisted activity. At level~4, AI-disclosure count is $23$\% higher ($\beta = .21^{***}$) and disclosure share $4.0$\% higher, while AI-assisted share is $15$\% higher and AI-assisted throughput $41$\% higher ($\beta = .34^{***}$). Participation metrics also strengthen, with active contributors $19$\% higher ($\beta = .17^{***}$), core developers $17$\% higher ($\beta = .16^{**}$), and the maintainer-responded share $8.0$\% higher. Level~3 show the same direction, with disclosure count $13$\% higher ($\beta = .12^{***}$) and active contributors $17$\% higher ($\beta = .15^{***}$).

\subsubsection*{\dR~strengthens both review engagement and code quality at level~4}
Covering 230 policies that mandate accountability, level~4 is associated with $28$\% higher throughput ($\beta = .25^{**}$), $7$\% higher disclosure count ($\beta = .07^{***}$), $5$\% more reviewers per PR ($\beta = .05^{***}$) , and $9.0$\% higher bot first-response share. This level also yields higher code-quality improvement, as code smells decline $45$\% ($\beta = -.61^{**}$), cognitive complexity $47$\%  ($\beta = -.64^{*}$), and ncloc $37$\% ($\beta = -.47^{**}$).

\subsubsection*{\dA~show mixed signals, with quality gains offset by higher vulnerabilities}
With 260 of 385 policies at level~1, only level~3 (104) provides sufficient observations for estimation. At this level, contributor counts are lower (active contributors $-8$\%, $\beta = -.09^{***}$) and several quality metrics are better (code smells $-46$\%, cognitive complexity $-53$\%, ncloc $-44$\%), yet vulnerabilities are $22$\% higher ($\beta = .20^{**}$).

\subsubsection*{\dC~curbs AI-assisted throughput and disclosure in proportion to restriction severity}
When the policy restricts what AI may be used for, AI-usage and disclosure estimates diminish monotonically. Constraint levels~2--4 are associated with $16.5$\% ($\beta = -.18^{***}$) to $33.0$\% ($\beta = -.40^{***}$) lower AI-assisted throughput, $4.9$\% ($\beta = -.05^{**}$) to $17.3$\% ($\beta = -.19^{***}$) lower AI-disclosure count and up to $8.0$\% lower disclosure share at level~3, and $14.8$\% lower core-developer count ($\beta = -.16^{***}$) at level~3.

\subsubsection*{\dE~reduces vulnerabilities and accelerates maintainer response, at the cost of fewer active contributors.}
The clearest Enforcement signal is on safety. At levels~3 and~4, vulnerabilities are $8.6$\% ($\beta = -.09^{*}$) and $14.8$\% lower ($\beta = -.16^{*}$), respectively, and bot first-response share decreases $12$\% at level~3. Higher enforcement is also associated with fewer active contributors ($9.5$\% at level~4, $\beta = -.10^{**}$) and with a $9.0$\% higher maintainer-responded share at level~4.

\begin{observation}{4}
The dimension \dT~increases disclosure, AI-assisted activity, as well as community participation, with effects stronger from level~3 to~4.
The dimension \dR~strengthens both review engagement and code quality.
The dimension \dA~shows mixed results as quality metrics improve but vulnerabilities rise.
\dC~curbs AI-assisted throughput and disclosure, with stronger effects at higher levels.
\dE~reduces vulnerabilities and speeds up maintainer response at the cost of fewer active contributors.
\end{observation}

\begin{table}[!t]
\centering
\caption{\textbf{Model III: Heterogeneous effects by policy family}}
\label{tab:ddd}
\small
\setlength{\tabcolsep}{3pt}
\renewcommand{\arraystretch}{1.0}
\begin{tabularx}{\columnwidth}{@{}>{\raggedright\arraybackslash\hangindent=.2in}X *{5}{c}@{}}
\toprule
Outcome & Perm. & Disc. & Reg. & Quiet & Strict \\
\midrule
\multicolumn{6}{@{}l}{\textcolor{accentSPACE}{\textbf{\icSat~Satisfaction}}} \\
\midrule
Core developer count (log) & \textcolor{coefPos}{.15\sigC} & \textcolor{coefNS}{--} & \textcolor{coefNS}{--} & \textcolor{coefPos}{.04\sigB} & \textcolor{coefPos}{.04\sigB} \\
LOC per reviewer (log) & \textcolor{coefNS}{--} & \textcolor{coefNS}{--} & \textcolor{coefNS}{--} & \textcolor{coefNS}{--} & \textcolor{coefNS}{--} \\
Maintainer responded PR share & \textcolor{coefNS}{--} & \textcolor{coefPos}{.08\sigA} & \textcolor{coefPos}{.07\sigA} & \textcolor{coefPos}{.09\sigB} & \textcolor{coefPos}{.09\sigA} \\
\midrule
\multicolumn{6}{@{}l}{\textcolor{accentSPACE}{\textbf{\icPerf~Performance}}} \\
\midrule
AI-assisted PR throughput (log) & \textcolor{coefPos}{.34\sigC} & \textcolor{coefPos}{.01\sigC} & \textcolor{coefPos}{.10\sigC} & \textcolor{coefNeg}{-.13\sigC} & \textcolor{coefNeg}{-.04\sigC} \\
Large PR share & \textcolor{coefNeg}{-.03\sigC} & \textcolor{coefNS}{--} & \textcolor{coefNS}{--} & \textcolor{coefNS}{--} & \textcolor{coefNS}{--} \\
Duplicated lines density & \textcolor{coefNS}{--} & \textcolor{coefNS}{--} & \textcolor{coefNS}{--} & \textcolor{coefNS}{--} & \textcolor{coefNS}{--} \\
Vulnerabilities (log) & \textcolor{coefNS}{--} & \textcolor{coefPos}{.19\sigA} & \textcolor{coefNeg}{-.33\sigC} & \textcolor{coefNS}{--} & \textcolor{coefNS}{--} \\
Code smells (log) & \textcolor{coefNS}{--} & \textcolor{coefNS}{--} & \textcolor{coefNS}{--} & \textcolor{coefNS}{--} & \textcolor{coefNeg}{-.41\sigA} \\
Cognitive complexity (log) & \textcolor{coefNS}{--} & \textcolor{coefNS}{--} & \textcolor{coefNS}{--} & \textcolor{coefNS}{--} & \textcolor{coefNS}{--} \\
\midrule
\multicolumn{6}{@{}l}{\textcolor{accentSPACE}{\textbf{\icAct~Activity}}} \\
\midrule
Active contributors (log) & \textcolor{coefPos}{.14\sigC} & \textcolor{coefPos}{.30\sigC} & \textcolor{coefNS}{--} & \textcolor{coefNS}{--} & \textcolor{coefPos}{.06\sigA} \\
AI-assisted PR share & \textcolor{coefPos}{.13\sigC} & \textcolor{coefPos}{.04\sigC} & \textcolor{coefPos}{.09\sigA} & \textcolor{coefNeg}{-.03\sigC} & \textcolor{coefPos}{.04\sigC} \\
Lines of code (ncloc) (log) & \textcolor{coefNS}{--} & \textcolor{coefNS}{--} & \textcolor{coefNeg}{-.29\sigB} & \textcolor{coefNeg}{-.34\sigB} & \textcolor{coefNeg}{-.45\sigB} \\
\midrule
\multicolumn{6}{@{}l}{\textcolor{accentSPACE}{\textbf{\icComm~Communication}}} \\
\midrule
AI disclosure PR count (log) & \textcolor{coefPos}{.17\sigC} & \textcolor{coefPos}{.10\sigB} & \textcolor{coefNS}{--} & \textcolor{coefNeg}{-.03\sigC} & \textcolor{coefPos}{.03\sigC} \\
AI disclosure PR share & \textcolor{coefPos}{.05\sigC} & \textcolor{coefPos}{.03\sigC} & \textcolor{coefPos}{.07\sigC} & \textcolor{coefNeg}{-.00\sigC} & \textcolor{coefPos}{.03\sigB} \\
Non-bot comments/PR (log) & \textcolor{coefNS}{--} & \textcolor{coefNS}{--} & \textcolor{coefNS}{--} & \textcolor{coefNS}{--} & \textcolor{coefNS}{--} \\
Bot first response share & \textcolor{coefNS}{--} & \textcolor{coefNS}{--} & \textcolor{coefNeg}{-.07\sigB} & \textcolor{coefNeg}{-.06\sigA} & \textcolor{coefNS}{--} \\
\midrule
\multicolumn{6}{@{}l}{\textcolor{accentSPACE}{\textbf{\icEff~Efficiency}}} \\
\midrule
Review turns/PR (log) & \textcolor{coefPos}{.19\sigA} & \textcolor{coefNS}{--} & \textcolor{coefNS}{--} & \textcolor{coefNS}{--} & \textcolor{coefNS}{--} \\
Reviewers/PR (log) & \textcolor{coefNS}{--} & \textcolor{coefNS}{--} & \textcolor{coefPos}{.07\sigA} & \textcolor{coefNS}{--} & \textcolor{coefNeg}{-.04\sigA} \\
Maintainer response time (log) & \textcolor{coefNS}{--} & \textcolor{coefNS}{--} & \textcolor{coefNS}{--} & \textcolor{coefNS}{--} & \textcolor{coefNS}{--} \\
\bottomrule
\end{tabularx}

\par\medskip
Each cell is the ATT coefficient $\beta$ against the matched control cohort with Permissive \& Open Use as referenced family; \\
``\textcolor{coefNS}{--}'' marks $p\ge0.05$;  $^{***}p<0.001$, $^{**}p<0.01$, $^{*}p<0.05$.
\\ Column headers abbreviate family names: Permissive \& Open; Disclosed Soft-Enforced; Regulated \& Controlled; Quiet \& Low-Requirement; Strict Prohibition.
\end{table}

\subsection{Heterogeneous Effects by Different Policy Families}

Table~\ref{tab:ddd} reports heterogeneous results by different policy families. Permissive \& Open Use is the reference family , so a starred cell differs from this baseline, while displayed coefficients reconstruct each family's absolute ATT against matched controls.

\subsubsection{\textbf{Permissive \& Open} policies yield the largest AI-assisted activity gains} AI-assisted share is $13$\% higher and AI-assisted throughput $40$\% higher ($\beta = .34^{***}$), while disclosure count is $18$\% higher ($\beta = .17^{***}$), active contributors $15$\% higher ($\beta = .14^{***}$), and core developers $16$\% higher ($\beta = .15^{***}$).

\subsubsection{\textbf{Disclosed but Soft-Enforced} policies produce the largest active-contributor gains, but also higher vulnerabilities} Active contributors are $34$\% higher ($\beta = .30^{***}$) and disclosure count $11$\% higher ($\beta = .10^{**}$). Vulnerabilities are, however,  $21$\% higher ($\beta = .19^{*}$).

\subsubsection{\textbf{Regulated \& Controlled} policies combine disclosure with stronger review and lower vulnerabilities} Disclosure share is $7.0$\% higher and reviewers $7.3$\% higher ($\beta = .07^{*}$), while vulnerabilities are $28$\% lower ($\beta = -.33^{***}$). AI-assisted share ($9.0$\% higher) and throughput ($11$\% higher, $\beta = .10^{***}$) also increase.

\subsubsection{\textbf{Quiet \& Low-Requirement}  policies are the only family with a net reduction in AI-assisted activity} AI-assisted merged throughput is $12$\% lower and AI-assisted share $3.0$\% lower, with disclosure count also $3.0$\% lower ($\beta = -.03^{***}$).

\subsubsection{\textbf{Strict Prohibition}  policies reduce AI throughput but do not eliminate observed AI-assisted activity} AI-assisted merged throughput is $3.9$\% lower ($\beta = -.04^{***}$), yet AI-assisted share ($4.0$\% higher) and disclosure share ($3.0$\% higher) remain positive. The strongest effects appear in quality and engagement as code smells are $34$\% lower, $ncloc$ 36\% lower, and the maintainer-responded share $9.0$\% higher.

\begin{observation}{5}
\textbf{Permissive} policies yield the largest AI-assisted activity gains.
\textbf{Disclosed-Soft} policies produce the largest contributor increase, but with higher vulnerabilities.
\textbf{Regulated} policies uniquely combine disclosure with stronger review and fewer vulnerabilities.
\textbf{Quiet} policies are the only family showing a net reduction in AI-assisted activity.
\textbf{Strict Prohibition} reduces AI throughput but does not eliminate observed AI use, and shows the largest quality improvements.
\end{observation}

\section{Threats to Validity}
\paragraph*{Internal Validity}
DID validity relies on the parallel-trend assumption, which we supported by event-study regressions where all 19 outcomes passed the pre-trend test (weeks -8 to -2). Robustness checks covering 4, 6, 10, and 12-week windows yielded directionally consistent results, suggesting reported short-term effects are not artifacts of window choice.
\paragraph*{External Validity}
AI policy adoption is recent and still accelerating (Fig.~\ref{fig:adoption}). Our post-adoption window setting is based on the objective constraint rather than a methodological preference. Whether early effect changes remains an open question until adopted policies age sufficiently.
\paragraph*{Construct Validity}
We define treatment as the initial adoption of a human-facing AI policy. Consequently, we excluded agent-guidance configurations (\texttt{AGENTS.md}) to avoid unmodeled, entangled effects. Besides, since repositories may revise policy when new frontier models launch, the effect of policy evolution needs more exploration.
\paragraph*{Conclusion Validity}  Data sparsity limits statistical power at certain TRACE levels containing fewer than 20 treated policies (T2: 4, A2: 7, E2: 8, A4: 14). We leave these cells unmarked as unreported, leaving the impact of under-represented configurations unknown.

\section{Discussion and Conclusion}
We now discuss the key findings and directions for practitioners and further research.
\label{sec:insight}

\subsection{AI governance makes AI-assisted work visible rather than banning it, and most OSS projects have not yet acted.}

\textbf{Policies function as signals that reduce uncertainty and encourage participation.} Only 1.3\% (385 of 29,624) sampled projects adopted an explicit AI policy. Yet for adopters, formal rules  legitimize AI use rather than suppress it as  AI-assisted activity, disclosure, and contributor counts all rise together. Higher \dT levels reinforce this, matching prior evidence that explicit norms reduce friction in OSS \cite{sun2026beyond, coelho2017modern}.

\textbf{Policies reshape the code review process toward lighter, more interactive cycles.} Post-adoption, contributions become more modular and reviews shift into a more dialogic loop. The \dR and \dE dimensions drive code review to a frequent, iterative human verification process.

\textbf{Strict rules raise internal quality standards rather than deterring participation.} Prohibition filters rather than forbids. Even under Strict Prohibition, AI use persists while quality improves most. Bans raise the entry bar for contributions rather than eliminate AI. Pairing \dT with \dE can lower vulnerabilities while sustaining throughput.

\subsection{OSS projects should prioritize transparency and balanced oversight rather than strict prohibition or opaque restrictions.}
\textbf{Make rules explicit rather than silent.} Silence carries a cost. Projects should state expectations clearly. Opaque restriction is the only configuration in our data associated with net reductions in AI activity and engagement. Stating nothing is worse than an explicit stance statement.

\textbf{Pair disclosure requirements with review mechanisms.} Requiring disclosure without review may attract contributors but introduce quality risks. Projects should consider combining visibility mandates (T$\geq$3) with validation requirements (R4) and credible enforcement (E$\geq$3) to convert transparency into accountability.

\textbf{Set enforcement and prohibition carefully.} Blanket bans are difficult to enforce and unnecessary. Moderate constraint (C2 or C3) paired with high transparency and appropriate enforcement can filter low-quality AI contributions while preserving the productivity and engagement benefits AI offers.

\subsection{Using TRACE as a reusable framework for future OSS AI governance research.}

Each TRACE dimension captures a distinct mechanism (\dT drives disclosure, \dR activates review quality, \dC modulates AI volume, \dE shapes security outcomes). This distinctness lets TRACE serve as a reusable tool for future OSS AI governance research. As AI-assisted contributions scale, trust hinges on verifiable and human-accountable code; the communities best positioned to supply this information are those making AI use visible and reviewable. Some areas where we expect TRACE to be reused are as follows:

\begin{itemize}
    \item \emph{As a foundation for predictive governance models.} Future work can use TRACE to build predictive models of how governance choices will shape community behavior under new AI capabilities.
    \item \emph{As a lens for studying policy evolution.} As AI tools evolve, OSS policies will not remain static. TRACE provides a stable vocabulary for measuring how policies tighten or relax over time, enabling future work to study governance evolution as a dynamic process.
    \item \emph{As a comparative framework across that enables cross-domain studies} of how transparency, responsibility, and enforcement shape quality and participation in different ecosystems (e.g., OSS vs.\ enterprise AI governance).
\end{itemize}

\subsection{Future work.}
Several open questions follow. First, we examine what policies contain but not how OSS communities form them, so future work can trace how AI policies emerge, from open contributor deliberation or top-down maintainer mandates, and then distinguish related legitimacy and efficacy. Second, current OSS governance remains largely reactive to frontier model releases, making anticipatory governance a promising direction. Finally, AI policies do not exist in isolation, and studying how they entangle with existing artifacts such as Codes of Conduct, governance structures, and agent-guidance files (\texttt{AGENTS.md}) would clarify whether AI policy is a  community norm or merely an isolated regulatory layer.

\bibliographystyle{IEEEtran}
\bibliography{references}

\end{document}